\documentclass[%
reprint,
superscriptaddress,
 amsmath,amssymb,
 aps,
 prb,
]{revtex4-2}

\usepackage[colorlinks, linkcolor=blue, anchorcolor=blue, citecolor=blue, urlcolor=blue]{hyperref} 
\usepackage{graphicx}
\usepackage{dcolumn}
\usepackage{bm}
\usepackage{float}
\usepackage{multirow}
\usepackage{setspace}

\begin{document}

\title{Electrically tunable superconducting dimensionality in stripe-ordered monolayer MoTe$_2$}

\author{Tianyu~Qiu}
\thanks{These authors contributed equally to this work}
\affiliation{National Laboratory of Solid State Microstructures and Department of Physics, Nanjing University, Nanjing 210093, China.}

\author{Ziyi~Han}
\thanks{These authors contributed equally to this work}
\affiliation{Center for Advanced Quantum Studies, School of Physics and Astronomy, Institute for Advanced Study, Beijing Normal University, Beijing 100875, China.}

\author{Fei~Yang}
\thanks{These authors contributed equally to this work}
\affiliation{Department of Physics, The Hong Kong University of Science and Technology, Clear Water Bay, Kowloon, Hong Kong SAR.}

\author{Keyu~Wei}
\thanks{These authors contributed equally to this work}
\affiliation{National Laboratory of Solid State Microstructures and Department of Physics, Nanjing University, Nanjing 210093, China.}

\author{Liguo~Dou}
\affiliation{National Laboratory of Solid State Microstructures and Department of Physics, Nanjing University, Nanjing 210093, China.}

\author{Yong~Zhang}
\affiliation{Jiangsu Provincial Key Laboratory of Advanced Photonic and Electronic Materials, School of Electronic Science and Engineering, Nanjing University, Nanjing 210093, China.}

\author{Dekun Yang}
\affiliation{Department of Physics, The Pennsylvania State University, University Park, PA, 16802, USA.}
\affiliation{Clare College, University of Cambridge, Trinity Lane, Cambridge CB2 1TL, UK}

\author{Kenji~Watanabe}
\affiliation{Research Center for Electronic and Optical Materials, National Institute for Materials Science, 1-1 Namiki, Tsukuba 305-0044, Japan.}

\author{Takashi~Taniguchi}
\affiliation{Research Center for Materials Nanoarchitectonics, National Institute for Materials Science, 1-1 Namiki, Tsukuba 305-0044, Japan}

\author{Xuefeng~Wang}
\affiliation{Jiangsu Provincial Key Laboratory of Advanced Photonic and Electronic Materials, School of Electronic Science and Engineering, Nanjing University, Nanjing 210093, China.}
\affiliation{Collaborative Innovation Center of Advanced Microstructures, Nanjing University, Nanjing 210093, China.}

\author{Fucong~Fei}
\affiliation{National Laboratory of Solid State Microstructures and Department of Physics, Nanjing University, Nanjing 210093, China.}

\author{Fengqi~Song}
\affiliation{National Laboratory of Solid State Microstructures and Department of Physics, Nanjing University, Nanjing 210093, China.}
\affiliation{Collaborative Innovation Center of Advanced Microstructures, Nanjing University, Nanjing 210093, China.}
\affiliation{Jiangsu Physical Science Research Center, Nanjing 210093, China.}

\author{Yanyan~Shangguan}
\affiliation{National Laboratory of Solid State Microstructures and Department of Physics, Nanjing University, Nanjing 210093, China.}

\author{Shaochun~Li}
\author{Jinsheng~Wen}
\affiliation{National Laboratory of Solid State Microstructures and Department of Physics, Nanjing University, Nanjing 210093, China.}
\affiliation{Collaborative Innovation Center of Advanced Microstructures, Nanjing University, Nanjing 210093, China.}
\affiliation{Jiangsu Physical Science Research Center, Nanjing 210093, China.}

\author{Hong~Yao}
\affiliation{Institute for Advanced Study, Tsinghua University, Beijing 100084, China.}

\author{Chaoxing~Liu}
\affiliation{Department of Physics, The Pennsylvania State University, University Park, PA, 16802, USA.}

\author{Binghai~Yan}
\email{binghai.yan@psu.edu}
\affiliation{Department of Physics, The Pennsylvania State University, University Park, PA, 16802, USA.}

\author{Lin~He}
\email{helin@bnu.edu.cn}
\affiliation{Center for Advanced Quantum Studies, School of Physics and Astronomy, Institute for Advanced Study, Beijing Normal University, Beijing 100875, China.}

\author{Xiaoxiang~Xi}
\email{xxi@nju.edu.cn}
\affiliation{National Laboratory of Solid State Microstructures and Department of Physics, Nanjing University, Nanjing 210093, China.}
\affiliation{Collaborative Innovation Center of Advanced Microstructures, Nanjing University, Nanjing 210093, China.}
\affiliation{Jiangsu Physical Science Research Center, Nanjing 210093, China.}

\begin{abstract}
Dimensionality governs the stability of collective quantum states, including whether Cooper pairs can establish the phase coherence required for superconductivity. In solids, however, electronic dimensionality is usually encoded in the crystal structure and is difficult to control reversibly. Here we show evidence that monolayer 1$T^{\prime}$-MoTe$_2$ develops a stripe charge-density wave that reorganizes its low-energy electronic states into quasi-one-dimensional channels. Raman spectroscopy and scanning tunneling microscopy identify the stripe order, whereas directional transport and tunneling spectroscopy reveal superconducting correlations confined to the stripe direction and insulating behavior perpendicular to it. Electrostatic doping weakens the transverse insulating response and drives the system into phase-coherent two-dimensional superconductivity. These results show that stripe charge order can reconfigure superconducting dimensionality and make it electrically tunable within a single crystalline layer. 
\end{abstract}

\maketitle

\begin{figure*}[t]
\centering
\includegraphics[width=\linewidth]{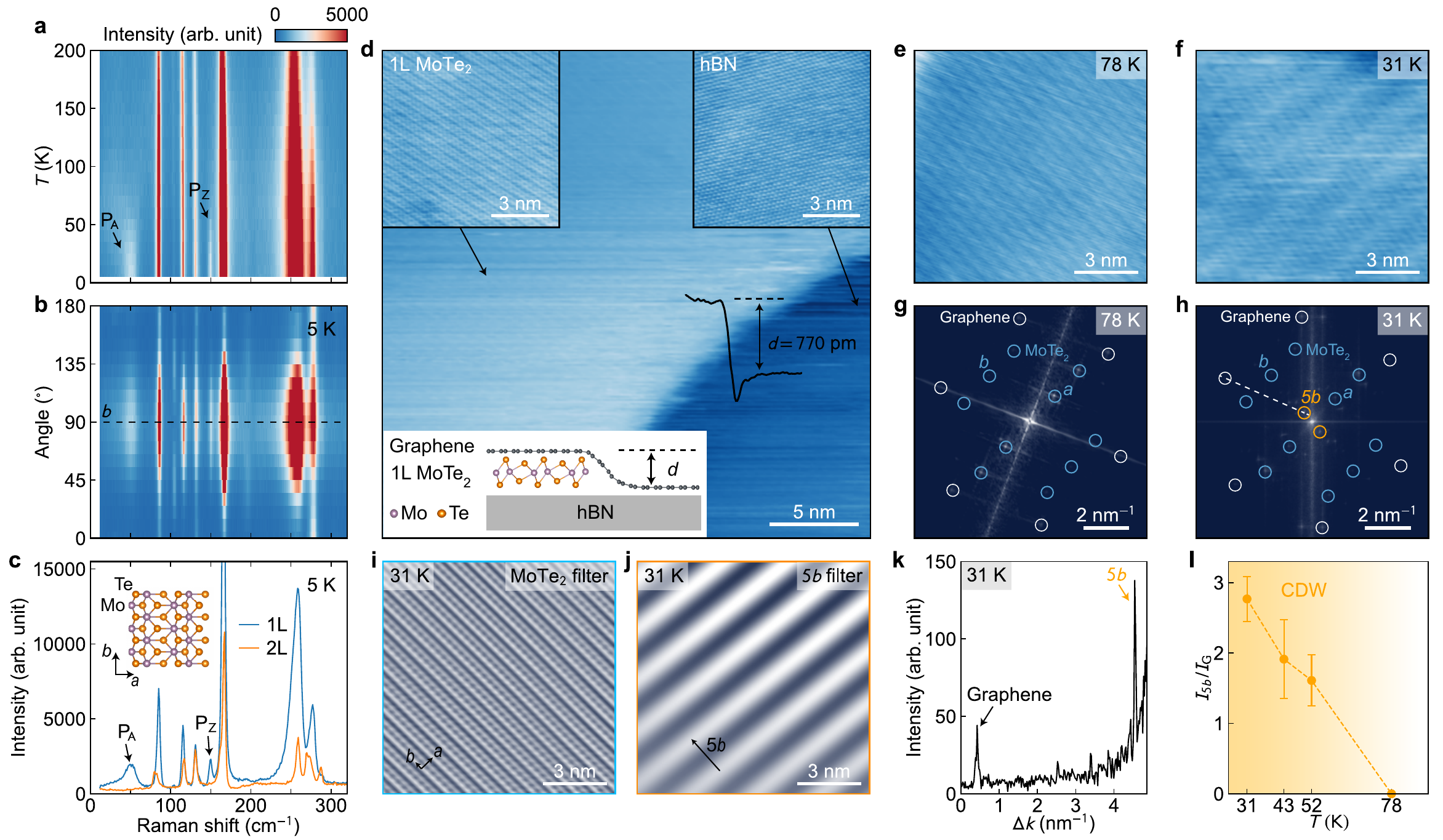}
\caption{\textbf{Signatures of stripe charge order.} \textbf{a} and \textbf{b}, Temperature- and polarization-angle-dependent Raman intensity maps of monolayer MoTe$_2$ (Sample M1). The dashed line in \textbf{b} marks the angle corresponding to the $b$ axis, where the $A_{\text{g}}$-symmetry modes reach maximum intensity. \textbf{c}, Raman spectra of monolayer and bilayer MoTe$_2$ at 5~K. Inset: monolayer crystal structure. Data in \textbf{a}--\textbf{c} were taken in parallel polarization configuration. \textbf{d}, STM topography around a tip-induced pit, with a height profile across the pit edge. Insets: topographies of monolayer MoTe$_2$ (Sample M2, top left) and the underlying hBN (top right), and a sample-structure schematic (bottom). \textbf{e} and \textbf{f}, STM topographies of monolayer MoTe$_2$ at 78~K and 31~K, respectively. Scanning parameters for \textbf{d}--\textbf{f} are given in Methods. \textbf{g} and \textbf{h}, FFT images of \textbf{e} and \textbf{f}, respectively. Blue, orange, and white circles mark the MoTe$_2$ reciprocal lattice, CDW superlattice, and graphene reciprocal lattice, respectively. \textbf{i} and \textbf{j}, Inverse FFT images of the blue- and orange-circled components in \textbf{h}, respectively. \textbf{k}, FFT profile along the dashed line in \textbf{h}, showing graphene ($I_{\text{G}}$) and CDW superlattice ($I_{5b}$) peaks. \textbf{l}, Temperature dependence of $I_{5b}/I_{\text{G}}$. }

\label{Fig1}
\end{figure*}

Superconductivity requires both Cooper pairing and global phase coherence. In reduced dimensions, enhanced fluctuations can separate the formation of Cooper pairs from the establishment of global phase coherence~\cite{Emery1995}. This separation is most pronounced in one dimension (1D), where superconducting correlations may develop but phase coherence is destroyed by fluctuations and phase slips~\cite{Bezryadin2000,Lau2001,Tian2005}. In quasi-1D systems consisting of an array of 1D channels, coupling neighboring channels can lock their phases and establish higher-dimensional superconductivity~\cite{Efetov1974,Emery1997,Carlson2000}. If the resulting state is effectively two dimensional (2D), phase coherence develops through the Berezinskii-Kosterlitz-Thouless (BKT) transition, where vortex-antivortex binding suppresses phase fluctuation driven by vortex motions~\cite{Berezinskii1971,Kosterlitz1973,Nelson1977}. This framework underlies studies of chain-like organic and inorganic superconductors and nanowire arrays~\cite{Lebed2008,Mitra2018,Yamaya2012,He2013}, but their quasi-1D character originates from pre-existing or engineered structures, limiting in-situ control of transverse coupling. Cold-atom systems have demonstrated controlled dimensional evolution between 1D correlation physics and higher-dimensional coherence~\cite{Vogler2014,Guo2024}, motivating the search for a tunable solid-state realization.

Stripe order offers a distinct route to controlling superconducting dimensionality. In stripe-ordered cuprates, charge and spin correlations are intertwined with superconductivity and can frustrate interlayer Josephson coupling~\cite{Tranquada1995,Himeda2002,Li2007,Berg2007}. More recently, stripe or smectic electronic phases have been found in a broader range of materials~\cite{Liu2021,Hua2024,Liu2023,Zhao2023,Devarakonda2024,Butler2026}, opening opportunities to explore how unidirectional electronic order affects superconductivity. In principle, stripe order can reconstruct a 2D electronic system into conducting channels separated by electronically depleted regions, suppressing transverse phase stiffness while preserving the underlying crystal structure. Direct evidence for such a stripe-induced quasi-1D superconducting regime and for its continuous tuning into phase-coherent 2D superconductivity has remained elusive.

Here we demonstrate this scenario in monolayer 1$T^{\prime}$-MoTe$_2$. Raman spectroscopy and scanning tunneling microscopy (STM) reveal a monolayer-specific stripe charge-density-wave (CDW) order that disappears on warming through about 80~K. In this stripe-ordered state, superconducting correlations develop along the stripes near an onset temperature ($T_c$) of 7~K, whereas charge transport perpendicular to the stripes becomes insulating and no robust BKT transition is observed, indicating insufficient transverse phase stiffness for global 2D superconducting coherence. Electrostatic doping suppresses the insulating response and drives the system from quasi-1D superconducting correlations into a phase-coherent 2D superconducting state, marked by zero resistance and clear BKT behaviour. These results establish monolayer MoTe$_2$ as a structurally simple and electrically tunable platform in which electronically driven stripe order reorganizes the dimensionality of superconductivity.

\begin{figure*}[t]
\centering
\includegraphics[width=0.8\linewidth]{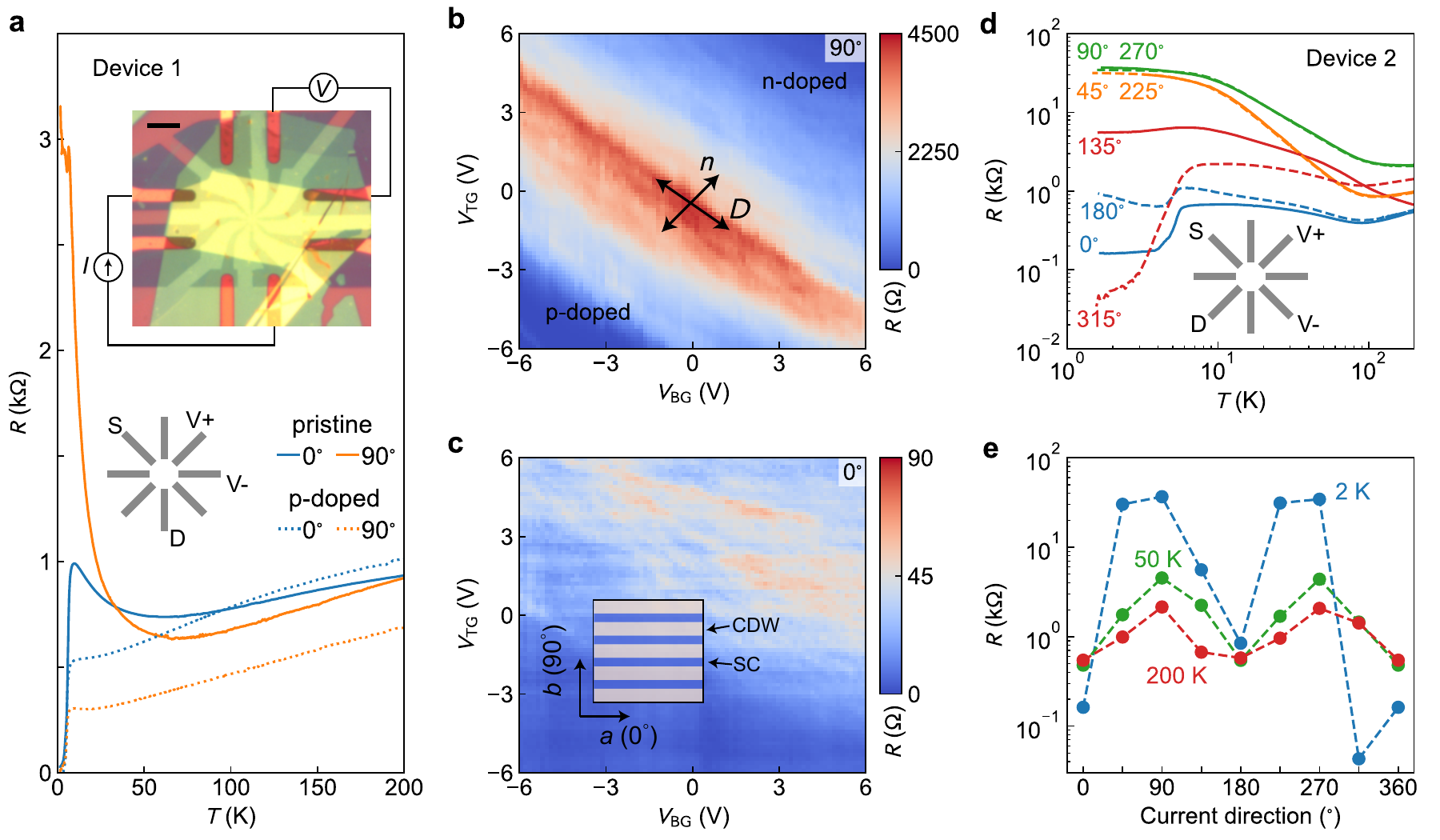}
\caption{\textbf{Anisotropic electrical transport.} \textbf{a}, Temperature-dependent resistance of Device~1 under pristine and hole-doped conditions, measured along two orthogonal directions. Upper inset: optical image of a typical device. The monolayer MoTe$_2$ is invisible due to its low contrast. $I$: Current source. $V$: Voltage probes. Scale bar: 10~$\mu$m. Lower inset: measurement scheme. S: Source. D: Drain. V$+$ and V$-$: Voltage probes. \textbf{b} and \textbf{c}, Resistance maps at 1.6~K as a function of the top- and bottom-gate voltages for the two directions shown in \textbf{a}. Arrows indicate the carrier-density ($n$) and displacement field ($D$) axes. Electron- and hole-doped regimes are labeled n-doped and p-doped, respectively. The inset in \textbf{c} illustrates the stripe CDW oriented along the $a$ axis. \textbf{d}, Temperature-dependent resistance of Device~2 measured along different directions using the scheme shown in the inset. \textbf{e}, Current-direction-dependent resistance at selected temperatures extracted from \textbf{d}.}
\label{Fig2}
\end{figure*}

\vspace{1mm}
\noindent
\textbf{Stripe charge order}

Bulk $T_d$-MoTe$_2$ is a type-II Weyl semimetal~\cite{Sun2015,Deng2016} that becomes superconducting near 0.1~K~\cite{Qi2016}. The ground state of monolayer MoTe$_2$ remains unsettled. Although it was predicted to be a quantum spin Hall insulator~\cite{Qian2014}, mechanically exfoliated samples have shown either strongly enhanced superconductivity~\cite{Rhodes2021} or a metal-insulator transition followed by superconducting onset~\cite{Tang2023}. Recent STM studies of epitaxial monolayers on graphene/SiC further reported stripe CDW, pair-density-wave (PDW) order, and intra-unit-cell superconducting gap modulation~\cite{Wei2025,Cheng2025b}. However, substrate-induced doping leaves the intrinsic ground state of pristine monolayer MoTe$_2$ unresolved.

We first establish that mechanically exfoliated monolayer MoTe$_2$ develops a stripe CDW order. Temperature- and polarization-dependent Raman measurements resolve the nine Raman-active phonon modes expected for the distorted $1T^{\prime}$ lattice~\cite{Marini2021,Lee2021,Cui2023}, together with two additional low-temperature modes, P$_{\text{A}}$ and P$_{\text{Z}}$, that appear only in the monolayer (Figs.~\ref{Fig1}a--c and Fig.~S1). Their temperature evolution identifies P$_{\text{A}}$ as a collective mode of the CDW order parameter and P$_{\text{Z}}$ as a zone-folded phonon associated with the CDW superlattice~\cite{Grasset2019,Liu2022}. The softening and broadening of P$_{\text{A}}$, together with the loss of P$_{\text{Z}}$ intensity upon warming, yield $T_{\mathrm{CDW}}\approx80$~K (Fig.~S2).

\begin{figure*}[t]
\centering
\includegraphics[width=\linewidth]{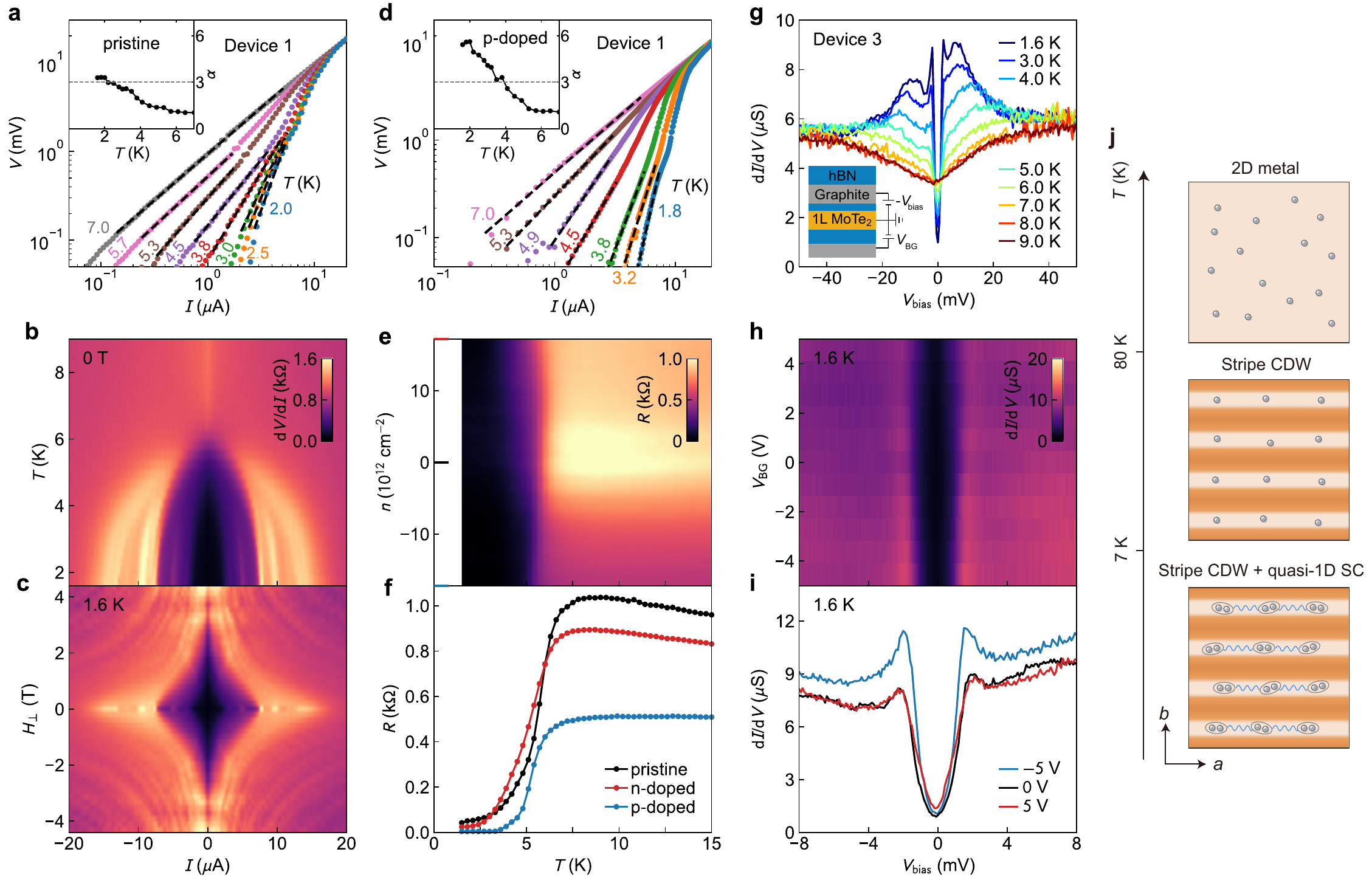}
\caption{\textbf{Quasi-1D superconducting correlations and gate-induced 2D coherence.} \textbf{a}--\textbf{f}, The d.c. transport data for Device~1. \textbf{a}, Temperature-dependent $I$-$V$ characteristics in the pristine state. Dashed lines are fits to $V\propto I^{\alpha}$; the extracted $\alpha$ is shown in the inset. \textbf{b} and \textbf{c}, The d.c.-current dependence of differential resistance at different temperatures (\textbf{b}) and out-of-plane magnetic fields (\textbf{c}). \textbf{d}, Temperature-dependent $I$-$V$ characteristics at maximum hole doping. \textbf{e}, Temperature- and doping-dependent resistance along the $a$ axis. \textbf{f}, Temperature line cuts from \textbf{e} in the pristine state and at maximum electron doping (n-doped) and hole doping (p-doped). \textbf{g}--\textbf{i}, Tunneling differential conductance data for Device~3. \textbf{g}, Temperature dependence. Inset: device-structure schematic. \textbf{h} and \textbf{i}, Gate-voltage-dependent conductance map at 1.6~K (\textbf{h}) and spectra at selected gate voltages (\textbf{i}). \textbf{j}, Schematic sequence of phase transitions. The stripe CDW suppresses transverse electronic motion and phase coupling, making superconducting correlations detectable only in transport along the $a$ axis.}
\label{Fig3}
\end{figure*}

STM measurements directly visualize the corresponding real-space modulation. We measured monolayer MoTe$_2$ encapsulated between hBN and a single-layer graphene capping layer (Fig.~\ref{Fig1}d), which preserves the CDW Raman signatures (Fig.~S3). At 78~K, STM topography and the corresponding FFT image show only the expected $1T^{\prime}$ lattice (Figs.~\ref{Fig1}e and g). At 31~K, a unidirectional modulation appears, producing additional FFT peaks corresponding to a periodicity of approximately $5b\approx17.5$~\AA~\cite{Keum2015} (Figs.~\ref{Fig1}f and h). Inverse FFT images with corresponding filters separately resolve the atomic lattice and the $5b$ modulation (Figs.~\ref{Fig1}i--j). The CDW FFT intensity, normalized by the weakly temperature-dependent graphene signal (Fig.~S4), decreases upon warming and vanishes by 78~K (Fig.~\ref{Fig1}k--l), consistent with the transition temperature inferred from Raman scattering. The modulation is absent in adjacent bilayer MoTe$_2$ regions (Fig.~S5), establishing a monolayer-specific unidirectional stripe CDW in pristine monolayer MoTe$_2$.

\vspace{1mm}
\noindent
\textbf{Anisotropic electrical transport}

Having established the stripe CDW order, we next examine its impact on in-plane charge transport. Device~1 consists of a dual-gated hBN-encapsulated monolayer MoTe$_2$, with radially arranged graphite contacts separated by $45^{\circ}$, allowing resistance measurements with electrical current passing along different in-plane directions (Fig.~\ref{Fig2}a). The $90^{\circ}$ direction is assigned to be approximately along the crystallographic $b$ axis using polarization-resolved Raman measurements (Fig.~\ref{Fig1}b)~\cite{Beams2016}. 

\begin{figure*}[t]
\centering
\includegraphics[width=0.8\linewidth]{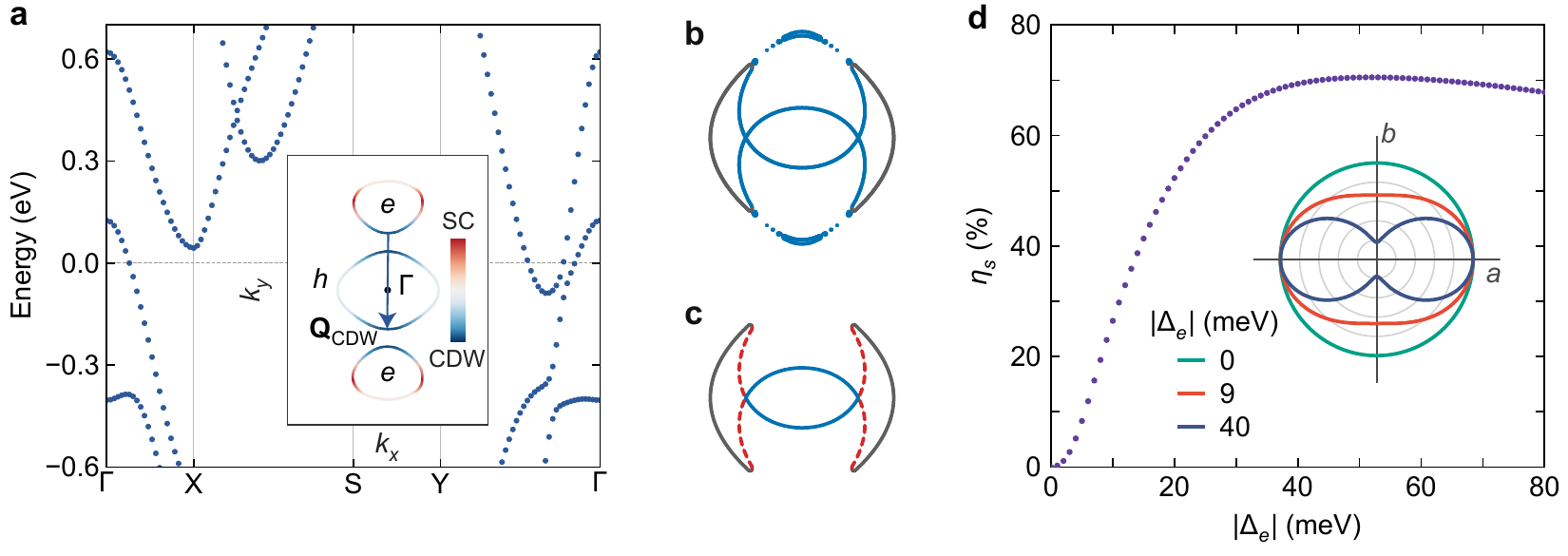}
\caption{\textbf{Theoretical calculations.} \textbf{a}, DFT band structure of monolayer MoTe$_2$ and corresponding Fermi surface (inset). $e$ and $h$ denote the electron and hole pockets, respectively. Colors schematically indicate the strength of CDW and superconducting correlations. The arrow marks the CDW ordering vector $\mathbf{Q}_{\rm CDW}$. \textbf{b} and \textbf{c}, Fermi surface folded by $\mathbf{Q}_{\rm CDW}$ before (\textbf{b}) and after (\textbf{c}) opening of the CDW gap $|\Delta_e|$. Dotted lines in \textbf{b} mark nested electronic states removed by the CDW gap. The remaining electron Fermi-surface segments with available opposite relative-momentum counterparts, highlighted by red dashed lines in \textbf{c}, support superconducting pairing. \textbf{d}, Superconducting anisotropy ratio as a function of $|\Delta_e|$. Inset: normalized angular-dependent superfluid stiffness $\rho_s(\theta)$ for selected values of $|\Delta_e|$.}
\label{Fig4}
\end{figure*}

The transport becomes strongly anisotropic in the CDW phase. At high temperature, resistance along the approximate $a$ and $b$ axes is metallic in both directions. Upon cooling into the stripe-ordered state, the $b$-axis resistance increases steeply, whereas the $a$-axis resistance shows only a weak upturn before dropping below about 7~K, signaling the onset of superconducting correlations (Fig.~\ref{Fig2}a). This indicates that the stripe CDW suppresses transport transverse to the stripes  while allowing superconducting correlations to develop along them. This directional dichotomy provides a natural explanation for previous reports of both superconducting~\cite{Rhodes2021} and insulating~\cite{Tang2023} low-temperature behavior in monolayer MoTe$_2$. By contrast, bilayer MoTe$_2$, which lacks the CDW order, remains metallic and superconducting along all measured directions (Fig.~S6).

Several checks confirm that this anisotropy is intrinsic to the stripe-ordered monolayer. Dual-gate maps at 1.6~K show that both electron and hole doping strongly suppress the insulating response along the $b$ axis, while the $a$-axis resistance is much less sensitive to doping (Figs.~\ref{Fig2}b--c). At the highest hole doping, transport becomes metallic in the normal state and superconducting at low temperature along both directions (Fig.~\ref{Fig2}a, dotted lines), showing that the extreme anisotropy is characteristic of the undoped stripe-CDW state. Device~2, measured using a different contact configuration, reproduces the same behavior and exhibits a clear two-fold symmetry, with $R_{90^{\circ}}/R_{0^{\circ}}$ exceeding 200 at 2~K (Figs.~\ref{Fig2}d--e). Similar behavior is observed in another device (Fig.~S7). Phenomenological simulations based on alternating superconducting and insulating stripes, as illustrated in Fig.~\ref{Fig2}c inset, qualitatively reproduce the direction-dependent transport (Supplementary Note~1).

\vspace{1mm}
\noindent
\textbf{Tunable superconducting dimensionality}

Low-temperature transport further shows that the stripe-ordered state hosts quasi-1D superconducting correlations. In the pristine state, the $a$-axis current-voltage ($I$-$V$)  characteristics reveal superconducting correlations below about 7~K, but without a fully developed BKT transition (Fig.~\ref{Fig3}a). The exponent $\alpha$ in $V\propto I^{\alpha}$ does not show the pronounced increase beyond 3 expected for robust BKT phase ordering~\cite{Beasley1979,Resnick1981} but instead saturates at low temperature (Fig.~\ref{Fig3}a inset), consistent with the absence of zero resistance (Fig.~\ref{Fig3}f). Differential-resistance ($\mathrm{d}V/\mathrm{d}I$) measurements confirm that this superconducting response persists up to about 7~K and survives under out-of-plane magnetic fields up to about 4~T (Figs.~\ref{Fig3}b--c). Together with the insulating $b$-axis transport, these results indicate that Cooper pairing develops along the stripe direction while transverse phase stiffness remains insufficient to establish global phase coherence.

Gate-induced doping weakens the stripe-imposed dimensional reduction and enhances transverse phase coherence. At the highest hole doping, the $a$-axis resistance approaches zero and a clear BKT transition emerges, with $\alpha$ increasing well above 3 upon cooling (Figs.~\ref{Fig3}d and f). The $b$-axis transport evolves in parallel, with doping suppressing the insulating response and reducing the low-temperature resistance nearly to zero (Fig.~S8). A Hall-bar device shows the same suppression of the pristine insulating state and the emergence of zero resistance on both the electron- and hole-doped sides (Fig.~S9). These results indicate that electrostatic doping weakens the stripe-imposed transverse decoupling and drives the system from quasi-1D superconducting correlations to phase-coherent 2D superconductivity.

Tunneling spectroscopy further probes the CDW and superconducting energy scales and their evolution with doping. In a monolayer MoTe$_2$/hBN/graphite tunnel junction, the normal-state differential conductance ($\mathrm{d}I/\mathrm{d}V$) exhibits a broad partial gap within $\pm50$~mV, consistent with the stripe-CDW state (Fig.~\ref{Fig3}g). Upon cooling below about 7~K, a narrow superconducting gap with sharp coherence peaks emerges at low bias, and the gap magnitude at base temperature is estimated to be 1.6~meV (Supplementary Note 2). Gate-dependent spectra at 1.6~K show that the zero-bias conductance and onset $T_c$ change only weakly with doping (Figs.~\ref{Fig3}h--i and Fig.~S10), consistent with the weak doping dependence of the $a$-axis onset $T_c$ in d.c. transport (Fig.~\ref{Fig3}e). Thus, electrostatic doping mainly restores transverse phase coherence rather than substantially changing the pairing scale along the stripe direction. These observations support a scenario in which the stripe CDW reconstructs the electronic states and suppresses transverse phase stiffness, allowing superconducting correlations to develop primarily along the stripe direction without establishing robust 2D phase coherence, as illustrated in Fig.~\ref{Fig3}j.

\vspace{1mm}
\noindent
\textbf{Electronic structure and mean-field calculations}

To understand the origin of the experimental findings, we first examine the momentum-space geometry of the electronic structure. Density-functional theory (DFT) calculations reveal a hole pocket centered at $\Gamma$ and two electron pockets located at ${\bf Q}_e=(0,\pm \frac{\pi}{3b})$ (Fig.~\ref{Fig4}{a}). The electron and hole pockets are partially nested by the experimentally observed CDW ordering vector, $\mathbf{Q}_{\rm CDW}=(0,\frac{2\pi}{5b})$ (Fig.~\ref{Fig4}{b}), which maximizes their overlap and the associated electron-hole scattering phase space. This indicates the CDW as an electron-hole instability that lowers the electronic free energy by gapping an extended manifold of nearly degenerate states connected by $\mathbf{Q}_{\rm CDW}$, while leaving much of the remaining electron Fermi surface intact (Fig.~\ref{Fig4}{c}). 

Guided by previous studies of monolayer MoTe$_2$~\cite{Cheng2025b}, we consider spin-singlet superconducting pairing within each electron pocket between states at $(\pm{\bf Q}_e+{\bf k},\uparrow)$ and $
(\pm{\bf Q}_e-{\bf k},\downarrow)$, while the hole pocket remains nonsuperconducting (Supplementary Note~3). When the CDW develops, partial Fermi-surface gapping strongly constrains the paired electronic states available in this intrapocket channel (Fig.~\ref{Fig4}c). On the upper electron pocket, for example, CDW correlations are concentrated near the lower vertex, where the nesting condition is best satisfied. Removing these states also suppresses the pairing channel involving their opposite relative-momentum counterparts along the $k_y$ direction, causing superconducting correlations to concentrate on the ungapped electron Fermi-surface segments along $k_x$ (Fig.~\ref{Fig4}a inset). The resulting superfluid stiffness tensor $\rho_{s,ij}$ is shown in Fig.~\ref{Fig4}{d} inset. The anisotropy ratio, $\eta_s=\frac{\rho_{s,xx}-\rho_{s,yy}}
{\rho_{s,xx}+\rho_{s,yy}}$, is nearly zero without CDW order and increases  with the CDW gap $|\Delta_e|$, reaching $\sim70\%$ for strong CDW order. 

\vspace{1mm}
\noindent
\textbf{Discussion}

The theoretical model captures the strong anisotropy but predicts phase-coherent 2D superconductivity in the undoped state, whereas the experiment reveals quasi-1D superconducting correlations without global phase coherence. This discrepancy suggests effects beyond the homogeneous mean-field treatment. In an anisotropic phase-only model, BKT fluctuations renormalize the superfluid stiffness but yield a single thermodynamic transition~\cite{Cai2026}, and therefore cannot explain the $b$-axis insulating response. Disorder provides a plausible mechanism. Te vacancies and fabrication-induced disorder (Supplementary Note~4) may randomize the transverse coupling between stripe-confined superconducting regions and suppress global phase coherence~\cite{Nakhmedov2010}. It remains to be seen whether cleaner undoped monolayers can spontaneously develop phase-coherent superconductivity. Further studies on gate-tuned devices may access the regime of extremely anisotropic 2D superconductivity~\cite{Huang2026,Xu2026} and clarify whether the evolution from quasi-1D correlations to 2D superconductivity is a crossover or a phase transition. 

Our results reveal an unusual coexistence of charge order and superconductivity, whose microscopic origin calls for momentum and spatially resolved probes across $T_c$. This issue is particularly important in view of the PDW state reported in epitaxial monolayer MoTe$_2$~\cite{Wei2025,Cheng2025b}, where evidence has so far come mainly from STM. The stripe-confined superconducting response observed here may provide a transport signature of unidirectional PDW order in atomically thin systems. The transport anisotropy observed here exceeds that of established related systems, where anisotropy either manifests as direction-dependent $T_c$ or is restricted to the normal state~\cite{Liu2021,Hua2024,Chu2010}. Tunneling spectra above $T_c$ in the CDW phase do not support a clean symmetric Luttinger-liquid regime (Supplementary Note~5), consistent with the dispersive reconstructed electronic structure (Fig.~\ref{Fig4}c). Thus, although the stripe CDW selectively confines superconducting correlations to the stripe direction, it does not realize a smectic metal above $T_c$~\cite{Emery2000}. This motivates controlled tuning or design of stripe-order-induced reconstruction that retains gapless quasi-1D channels, opening a route to non-Fermi-liquid phases in 2D systems~\cite{Vishwanath2001}.

\vspace{3mm}
\noindent
\textbf{Acknowledgments}\\
We thank Kin Fai Mak for helpful discussions. This work was supported by the National Key Research and Development Program of China (Grant No. 2024YFA1409100), the Natural Science Foundation of Jiangsu Province (Grants No. BK20231529 and No. BK20233001), the National Natural Science Foundation of China (Grants No. U24A6002, No. 12474170, No. 12425405, No. 62525406 and No. T2394473), the Fundamental Research Funds for the Central Universities (No. 0204-14380264), and the Innovation Program for Quantum Science and Technology of China (Grant No. 2024ZD0301300). K.W. and T.T. acknowledge support from the CREST (JPMJCR24A5), JST and World Premier International Research Center Initiative (WPI), MEXT, Japan.

\vspace{3mm}
\noindent
\textbf{Author contributions}\\
X.X. conceived and supervised the project. T.Q. and K.W. (NJU) prepared the atomically thin samples and fabricated the devices. T.Q., K.W. (NJU) and X.X. performed the Raman, transport and tunnelling spectroscopy measurements. Z.H. and L.H. performed the STM measurements on atomically thin samples. L.D. and S.L. performed the STM measurements on bulk crystals. Y.Z., X.W., F.F., F.S., Y.S. and J.W. grew the MoTe$_2$ crystals. K.W. (NIMS) and T.T. supplied the hBN crystals. T.Q., K.W. (NJU), Z.H., L.H. and X.X. analysed the experimental data. F.Y., D.Y. and B.Y. performed the theoretical calculations. X.X., C.L., B.Y. and H.Y. interpreted the results. X.X., F.Y. and B.Y. wrote the manuscript with input from all authors.

\vspace{3mm}
\noindent
\textbf{Competing interests}\\
The authors declare no competing interests.
\newpage

\end{document}